\documentstyle[12pt,aps,epsfig]{revtex}

\begin{document}

\preprint{KFA-IKP(TH)-1999-14 and NT@UW-99-31}

\title{What does ``$\rho$ exchange'' in $\pi N$ scattering mean?
}

\author{O. Krehl$^a$, C. Hanhart$^{b}$, S. Krewald$^a$, and J. Speth$^a$}

\address{$^a$ Institut f\"ur Kernphysik \\ 
Forschungszentrum J\"ulich GmbH, 52425 J\"ulich, 
Gemany\\
$^b$ Department of Physics and INT,\\
 University of Washington, Seattle, WA 98195, USA}

\date{June 23, 1999}

\maketitle

\vspace{-7cm} 
\hfill{KFA-IKP(TH)-1999-14 and NT@UW-99-31}
\vspace{7cm}

\begin{abstract}
We present an alternative method for calculating
amplitudes for correlated $\pi\pi$ exchange in the ``$\sigma$'' and $\rho$
channel in $\pi N$ scattering. Starting from a fixed mass meson exchange
potential, we introduce the width of the exchanged particles by integrating
over a mass spectral function. The spectral functions are constructed
from the pseudoempirical $N\bar{N}\to\pi\pi$ data. Using this
approach we develop a prescription for resolving ambiguities of the
correlated $\pi\pi$ exchange in the $\rho$ channel that occur when different
dispersion theoretical formulations of $\rho$ exchange are used to construct
$\pi N$ potentials.
\end{abstract}


\section{Introduction}

The exchange of meson pairs plays an important role in hadron
dynamics. The first meson exchange models of the two-nucleon
interaction were based on the exchange of one meson only and had to
introduce a scalar-isoscalar meson (the sigma) which is not seen as a
resonance in the two-pion phase shifts and has remained a
controversial issue \cite{PDG98}.  Models of the two-nucleon
interaction based on dispersion relations show that the sigma meson
can be understood as an effective degree of freedom which
parameterizes the exchange of two pions
\cite{Cott73,Brown76,Durso80,Machleidt87}.
A natural extension of this approach leads to the inclusion of $\pi\pi$ correlations
in the $\rho$-channel \cite{Reuber96}.
In addition to the exchange of two pions, the exchange of a pion-rho
pair is necessary to describe the phase shifts, in particular near the
pion production threshold \cite{Machleidt87}.

\begin{figure}[b]
\begin{center}
\epsfig{file=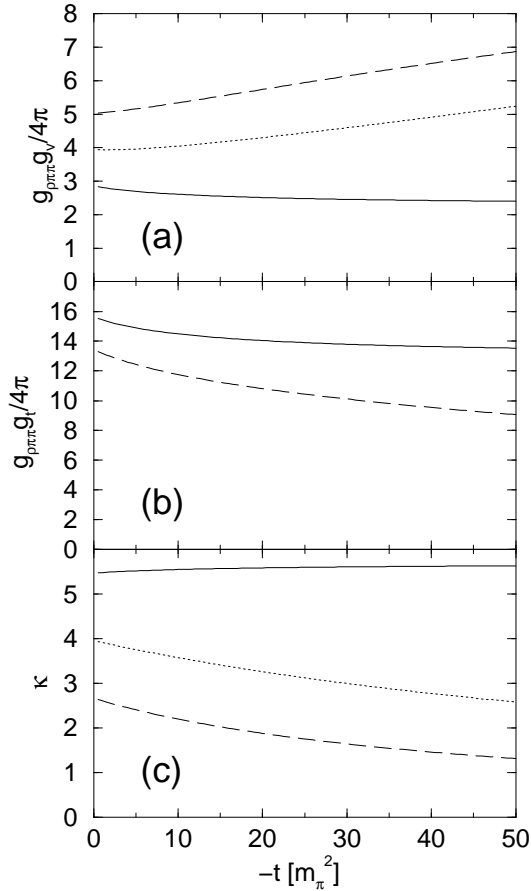,height=12.0cm}
\caption{Coupling constants $g_{\rho\pi\pi}g_v/4\pi$ (a),
$g_{\rho\pi\pi}g_t/4\pi$
(b) and the ratio
$\kappa=g_t/g_v$ (c) calculated by using different dispersion
relations as in Ref. \protect \cite{Schuetz95}.  The solid line is for the
$\Gamma$ amplitudes of eq. (\protect \ref{dispgamma}), the dashed line for the
$f$ amplitudes of eq. (\protect \ref{dispf}) and the dotted line for the
$h$ amplitude from eq. (\protect \ref{dispgh}). The $g$ amplitude is
proportional to the $\Gamma_2$ amplitude and therefore gives the same
tensor coupling as the $\Gamma$ amplitudes. The amplitudes are defined in
Sec. \protect \ref{secrho}}
\label{figcoupl}
\end{center}
\end{figure}

Correlated two-pion exchange has been incorporated in meson-theoretic
models of both pion-nucleon \cite{Schuetz94} and kaon-nucleon
scattering \cite{Hoffmann95}.  In these calculations ambiguities in
the dispersion theoretic approach \cite{Durso77} to the correlated
two-meson exchange in the $\rho$ channel were encountered \footnote{In
  the dispersion theoretic approach the $\rho$ exchange is defined by
  integrating right hand cuts only \cite{Hoehler75}. By leaving out
  contributions from the left hand cuts, the mentioned ambiguities
  arise.}. The quantitative effect of these ambiguities is illustrated
in Fig. \ref{figcoupl}, where we show the vector and tensor $\rho NN$
coupling as calculated using different dispersion relation
formulations \cite{Schuetz95}.  Given the important role of correlated
two-meson exchanges in many different hadronic reactions, one is
compelled to address the question which of the various dispersion
relation formulations is preferred.  In order to answer this question,
we calculate the correlated $\pi\pi$ exchange, not in the framework of
dispersion relations, which suffer from the problem of being
ambiguous, but in a formulation based on spectral functions.

We  introduce this method in the next section using the $\pi N$
potential due to correlated $\pi\pi$ exchange in the $\sigma$ channel as an
example, which is simpler than the exchange in the $\rho$ channel.  In
the third section we deal with the construction of the potential in
the $\rho$ channel and discuss the problem of differing
results in the dispersion relation approach.  The last section
summarizes our results.

\section{The $\pi N$ potential of correlated $\pi\pi$ exchange in the
$\sigma$ channel}\label{secsig}

\begin{figure}
\begin{center}
\epsfig{file=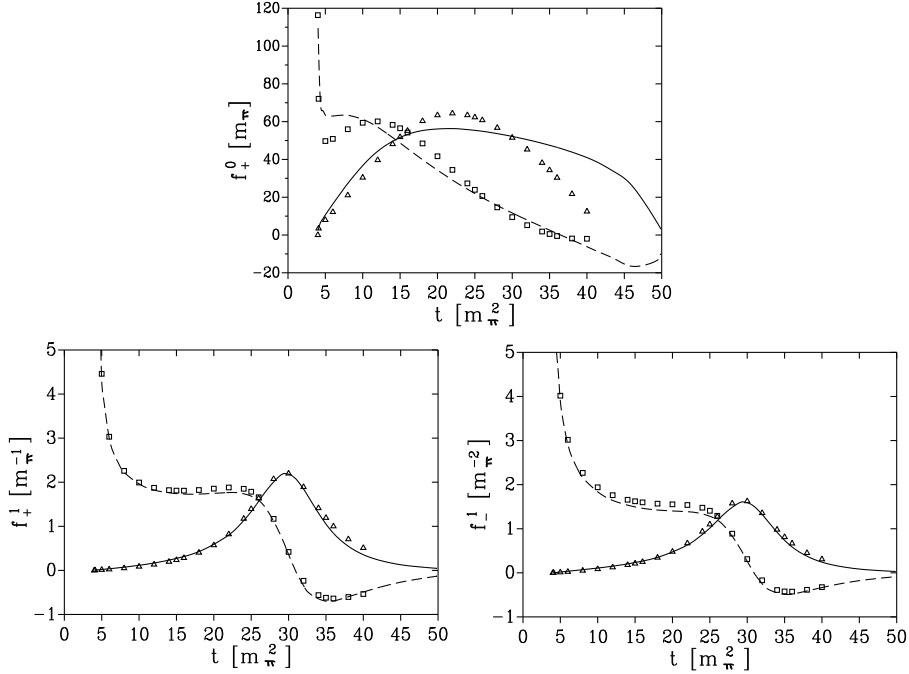,height=9cm}
\caption{Spectral functions ($Im(f)$) in the $\sigma$ ($f^0_+$) and
$\rho$ ($f^1_{\pm}$) channel of the reaction $N\bar{N} \to \pi\pi$. The
experimental
information is taken from Ref. \protect \cite{Hoehler83}. The curves are from
the
microscopical model of Ref. \protect \cite{Schuetz95}.
}
\label{figspec}
\end{center}
\end{figure}

In order to extract the dynamical information from the transfer
matrix in both channels, $N\bar{N}\to\pi\pi$ (hereafter called the
$t$-channel) and $\pi N \to \pi N$ (hereafter called the $s$-channel),
one must first decompose the $T$-matrix into well-defined
operators and amplitudes that contain the dynamics of the
reaction.  The most familiar amplitudes in this respect are the
$f$ amplitudes (see Fig. \ref{figspec}) introduced by Frazer and Fulco
\cite{Frazer60}, which we will use as input for our calculations.
The details of this decomposition are given in the appendix \ref{zerleg}.

Our starting point for the construction of the correlated $\pi\pi$ exchange
is the Lagrangian
\begin{equation}
{\cal L}_{int}=g_{NN\sigma}\bar{\Psi}\Psi\sigma +
\frac{g_{\sigma\pi\pi}}{2m_{\pi}}\partial_{\mu}\vec{\pi}
\cdot \partial^{\mu}\vec{\pi}\sigma.
\end{equation}
We use derivative coupling at the $\sigma\pi\pi$ vertex to ensure
chiral symmetry. From this Lagrangian we calculate a $\pi N$ potential
\begin{equation}
V_{\sigma}=\frac{g_{\sigma NN} g_{\sigma \pi\pi}}{2m_{\pi}}
2 p_{2_\mu} p^{\mu}_4  P_{\sigma}(t,m_{\sigma}^2)
\bar{u}(\vec{p}_3,\lambda_3)u(\vec{p}_1,\lambda_1),
\end{equation}
where the propagator $P_{\sigma}(t,m^2_{\sigma})$ depends on the theoretical
framework in which one is working. In a covariant approach the propagator
is
\begin{displaymath}
P_{\sigma}(t,m^2_{\sigma})=\frac{1}{t-m^2_{\sigma}}.
\end{displaymath}
Our notation is shown in Fig. \ref{fignot} in the appendix.

The potential $V_{\sigma}$ can be expressed in terms of the invariant
amplitudes $A$ and $B$ defined in the appendix:
\begin{eqnarray}\label{ampsigma}
A^{(+)}_{\sigma}&=&-\frac{g_{NN\sigma}g_{\sigma\pi\pi}}{2m_{\pi}}2p_{2_\mu}p^{\mu}_4
P_{\sigma}(t,m_{\sigma}^2),\nonumber \\
B^{(+)}_{\sigma}&=&0.
\end{eqnarray}

In order to obtain an idea of how to proceed, let us look again at the
spectral functions in Fig. \ref{figspec}. The two spectral functions
of the $\rho$ channel suggest a Breit-Wigner parameterization of
a resonance. The width of such a resonance is the
decay width of the particle. If the particle is stable the width is
zero, resulting in a delta function located at
the mass squared of the particle.  The spectral functions $(Im(f))$ can then
be regarded as mass distributions of an unstable particle. We can then
incorporate the width of the $\sigma$ meson into our amplitudes by
exchanging a continuum of $\sigma$ mesons with different masses
$m_{\sigma}$, with amplitudes weighted according to the spectral function.
This can be cast in spectral form \cite{Peters98,Rapp98,Rapp99}:
\begin{equation}\label{a2pi}
A^{(+)}_{2\pi}=\frac{1}{\pi}\int dm^2_{\sigma} \rho_{\sigma}(m^2_{\sigma})
A^{(+)}_{\sigma}(m^2_{\sigma}),
\end{equation}
where the spectral function is normalized according to
\begin{displaymath}
\frac{1}{\pi}\int dm^2_{\sigma} \rho_{\sigma}(m^2_{\sigma})=1.
\end{displaymath}

\begin{figure}
\begin{center}
\epsfig{file=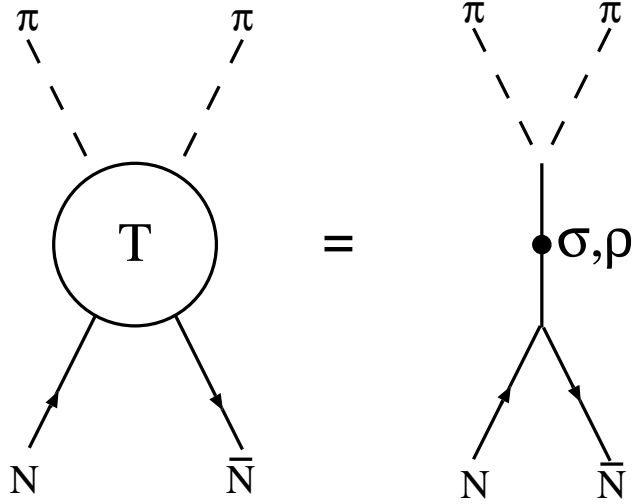,height=7cm}
\caption{Expressing the $N\bar{N}\to\pi\pi$ $T$-matrix in term of a meson pole
diagram with a dressed propagator, which contains all the dynamics.}
\label{figpic}
\end{center}
\end{figure}

All that we need do is calculate the spectral
function $\rho_{\sigma}(m_{\sigma})$.
We do this by relating the spectral function to the
dressed $\sigma$ propagator, $D_{\sigma}(m_{\sigma}^2)$, via
\begin{equation}\label{propspec}
\rho_{\sigma}(m^2_{\sigma})=-Im(D_{\sigma}(m^2_{\sigma})).
\end{equation}
In order to extract the $\sigma$ propagator, we express the
$N\bar{N} \to \pi\pi$ amplitudes of eq. (\ref{ampfsigma}) as
a $\sigma$ pole diagram with a dressed propagator, $D_{\sigma}(t)$,
which contains all the $\pi\pi$ dynamics, as shown in Fig. \ref{figpic}.
We thus obtain
\begin{equation}
A^{(+)}_{\sigma}=\frac{g_{\sigma NN} g_{\sigma \pi\pi}}{2m_{\pi}}2 q_{3_\mu}
q^{\mu}_4 D_{\sigma}(t)=
-\frac{4\pi}{p_t^2} f^0_+(t).
\end{equation}
The spectral function is then given by
\begin{equation}\label{sigmaspec}
\rho_{\sigma}(m^2_{\sigma})=-Im(D_{\sigma}(m^2_{\sigma}))=\frac{4\pi}{p_t^2}Im(f^0_+(t))
\frac{2m_{\pi}}{g_{\sigma\pi\pi}g_{\sigma NN}}\frac{1}{2 q_{3_\mu} q^{\mu}_4},
\end{equation}
from which we obtain the amplitude $A^{(+)}_{2\pi}$:
\begin{eqnarray}\label{ressigma}
A^{(+)}_{2\pi}&=&-4(2p_{2_\mu} p^{\mu}_4)\int dt'\frac{Im(f_+^0(t'))}
{p^2_{t'}(t'-2m^2_{\pi})}P_{\sigma}(t,t') \nonumber \\
&=&-16(2p_{2_\mu}  p^{\mu}_4)\int
dt'\frac{Im(f_+^0(t'))}{(t'-4m^2_N)(t'-2m^2_{\pi})} P_{\sigma}(t,t'),
\end{eqnarray}
Here we have used the relation $2 q_{3_\mu} q^{\mu}_4=t'-2m_{\pi}^2$
and the definition of the on shell momentum $p^2_{t'}=\frac{t'}{4}-m_N^2$ and
changed the variable of integration from $m^2_{\sigma}$ to $t'$. This
enables us to compare our expression directly with the results of the
dispersion theoretical approach of Ref. \cite{Schuetz94}
\begin{equation}\label{sigschuetz}
\tilde{A}^{(+)}_{2\pi}=16(t-2m_{\pi}^2)\int
dt'\frac{Im(f_+^0(t'))}{(t'-4m^2_N)(t-t')(t'-2m^2_{\pi})}.
\end{equation}
The tilde is used only as a reminder that this is a result from dispersion
theory.

\begin{figure}
\begin{center}
\epsfig{file=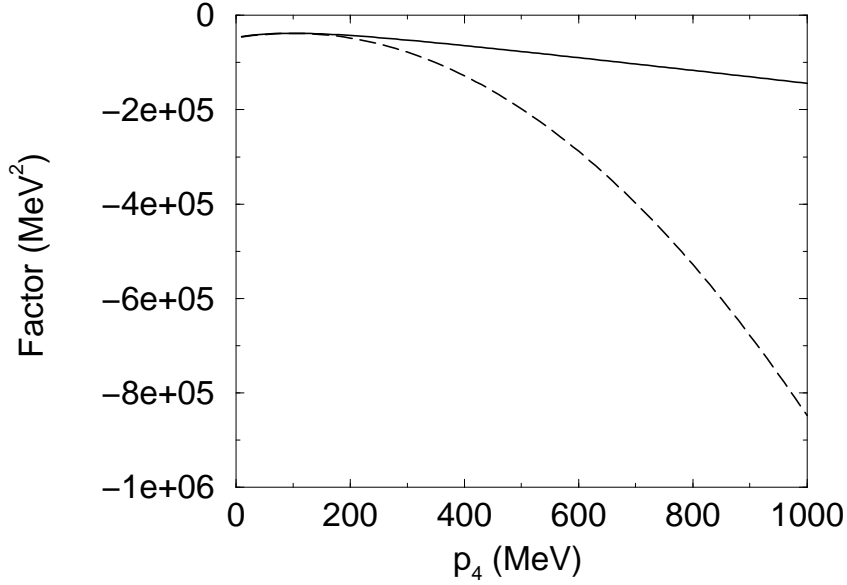,height=8cm}
\caption{The factors $(-2p_{2_\mu}p^{\mu}_4)$ (solid line)
and $(t-2m_{\pi}^2)$ (dotted line) as a function of the off-shell
momentum $|p_4|$ where $|p_2|$=100 MeV is fixed. The momenta $\vec{p}_2$ and
$\vec{p}_4$ are chosen to be parallel.} \label{figfaktor}
\end{center}
\end{figure}

In comparing the two results, eq. (\ref{ressigma}) derived by using
the spectral function and eq. (\ref{sigschuetz}) calculated within the
framework of dispersion relations, we find differences only in the
factor in front of the integral: $-(2p_{2_\mu}p^{\mu}_4)$ in eq.
(\ref{ressigma}) and $(t-2m_{\pi}^2)$ in eq. (\ref{sigschuetz}).  For
on-shell amplitudes, these two terms agree. Therefore the on-shell
amplitudes of both methods are the same, as it must be.
Off-shell, however, they
are very different, as seen in Fig. \ref{figfaktor}. If one wishes to
use this amplitude as a potential in a scattering equation or as an
off-shell $\pi N$ amplitude in another reaction, care must be taken
to use a potential with an off-shell behavior that is compatible with
the off-shell behavior of other diagrams (e.g. nucleon exchange).
The off-shell behavior can be implemented easily in eq.
(\ref{ressigma}), because here only the momenta of the two pions
appear and the off-shell prescription of these momenta is given by the formalism
used to solve the scattering equation 
for which the correlated $\pi\pi$ exchange serves as a potential.
The importance of a proper off-shell
behavior of the $\pi N$ amplitude constructed in a dispersion
theoretical way is also pointed out in the calculation of the $\pi NN$
vertex in Ref. \cite{Durso77}. There the authors prescribe the
off-shell behavior of the correlated two pion exchange by comparing
the amplitude to the field theoretical amplitude for the exchange of
a stable particle.

The fact that the on-shell amplitudes, which are the ones we are interested
in at the moment, are the same with both methods shows that our
method leads to correct results where dispersion theory is applicable.
This observation encourages us to proceed in the calculation of the
correlated $\pi\pi$ exchange in the $\rho$ channel.

\section{The correlated $\pi\pi$ exchange in the $\rho$ channel}\label{secrho}

Let us start this section by taking a closer look at the source of
the difference displayed in Fig. \ref{figcoupl}. The coupling constants
shown there are calculated by using different dispersion relations
\cite{Schuetz96J}.  The dashed lines were calculated by using the
amplitudes
\begin{eqnarray}\label{dispf}
\tilde{A}^{(-)}_{\rho}(s,t)&=&12\frac{p_tq_tx}{p^2_t}\left(\frac{m_N}{\sqrt{2}}\int
dt' \frac{Im(f^1_-(t'))}{t'-t}-\int dt' \frac{Im(f^1_+(t'))}{t'-t}\right)
\nonumber \\
\tilde{B}^{(-)}_{\rho}(s,t)&=&6\sqrt{2} \int dt' \frac{Im(f^1_-(t'))}{t'-t},
\end{eqnarray}
which we get by writing dispersion relations for $f^1_+$ and $f^1_-$
separately and inserting these into eq. (\ref{ampfrho}).

A different way was originally suggested by Frazer and Fulco
\cite{Frazer60} and later applied by H\"ohler and Pietarinen
\cite{Hoehler75} to a calculation of the the $\rho NN$ couplings.
They define the combinations
\begin{eqnarray}\label{gammaamp}
\Gamma_1(t)&=&-\frac{m_N}{p_t^2}
\left(f^1_+(t)-\frac{t}{4\sqrt{2}m_N}f^1_-(t)\right) \nonumber \\
\Gamma_2(t)&=&\frac{m_N}{p_t^2}
\left(f^1_+(t)-\frac{m_N}{\sqrt{2}}f^1_-(t)\right),
\end{eqnarray}
for which dispersion relation are written. This leads to the amplitudes
\begin{eqnarray}\label{dispgamma}
\tilde{A}^{(-)}_{\rho}(s,t)&=&-12\frac{p_tq_tx}{m_N} \int dt'
\frac{Im(\Gamma_2(t')}{t'-t} \nonumber \\
\tilde{B}^{(-)}_{\rho}(s,t)&=&12 \left(\int dt' \frac{Im(\Gamma_1(t'))}{t'-t} +
\int dt' \frac{Im(\Gamma_2(t'))}{t'-t}\right),
\end{eqnarray}
and to the solid curves in Fig. \ref{figcoupl}.
Note that the factor $\frac{1}{p_t^2}$ does not generate a pole
at $t=4m_N^2$ since the $f$ amplitudes obey the relation \cite{Frazer60}
\begin{equation}\label{cancel}
f^1_+(4m^2_N)-\frac{m_N}{\sqrt{2}}f^1_-(4m_N^2)=0.
\end{equation}

Yet another possibility was suggested by J.M. Richard \cite{Richard}. He
defines two amplitudes
\begin{eqnarray}
g(t)&=&f^1_+(t)-\frac{m_N}{\sqrt{2}} f^1_-(t) \nonumber \\
h(t)&=&f^1_+(t)+\frac{m_N}{\sqrt{2}} f^1_-(t),
\end{eqnarray}
and formulates a subtracted dispersion relation for $g$ and an
unsubtracted dispersion relation for $h$
\begin{eqnarray}\label{dispgh}
\tilde{g}(t)&=&\frac{t-4m_N^2}{\pi}\int dt' \frac{Im(g(t'))}{(t'-t)(t'-4m_N^2)}
\nonumber \\
\tilde{h}(t)&=&\frac{1}{\pi} \int dt' \frac{Im(h(t'))}{t'-t}.
\end{eqnarray}
Since $g(t)\propto\Gamma_2$ the amplitude $A^{(-)}_{2\pi}$ (and therefore the
tensor coupling)
is the same as in eq. \ref{dispgamma}. The amplitude $B^{(-)}_{2\pi}$, however,
reads now
\begin{equation}
\tilde{B}^{(-)}_{2\pi}=12\pi m_N(\tilde{h}(t)-\tilde{g}(t))
\end{equation}
and leads to a different vector coupling, shown in Fig. \ref{figcoupl}.

The fact that these three alternatives lead to very
different amplitudes can be seen directly by comparing the eqs.
(\ref{dispf}), (\ref{dispgamma}), and (\ref{dispgh}).  One can see for
example, that the factor $\frac{1}{p_t^2}$ appears in front of the
dispersion relation integral over the $f$ amplitudes in eq.
(\ref{dispf}), whereas it is absorbed in the definition of the
$\Gamma$ amplitudes in eq. (\ref{dispgamma}). Of course this
influences the convergence of the dispersion integrals which causes
the differences between the couplings displayed in Fig.
\ref{figcoupl}. These are the ambiguities mentioned above.

Let us now calculate the $\rho$ exchange by using the spectral
function approach we have introduced in section \ref{secsig}.
The calculation of the
amplitudes in the $\rho$ channel proceeds along the same lines as
outlined in the previous section.  The Lagrangian we start with is
\begin{equation}\label{lagrho}
{\cal L}_{int}=g_v\bar{\Psi}(\gamma_{\mu}\vec{\rho}^{\mu}-
\frac{\kappa}{2m_N}\sigma_{\mu \nu}\partial^{\nu}\vec{\rho}^{\mu})
\frac{\vec{\tau}}{2}\Psi +
g_{\rho\pi\pi}(\vec{\pi} \times \partial_{\mu} \vec{\pi}) \vec{\rho}^{\mu}
\end{equation}
($\kappa=\frac{g_t}{g_v}$).
From this we calculate the invariant amplitudes
\begin{eqnarray}\label{abrho}
A^{(-)}_{\rho}&=&g_{\rho\pi\pi}\frac{g_t}{m_N}\frac{1}{2}Q^{\mu}(p_1+p_3)_{\mu}
P_{\rho}(t,m^2_{\rho}) \nonumber \\
B^{(-)}_{\rho}&=&-g_{\rho\pi\pi}(g_v+g_t)P_{\rho}(t,m^2_{\rho}),
\end{eqnarray}
which we have to weight with a spectral function to get the correlated
$\pi\pi$ exchange.

The two coupling schemes of the $\rho$ to the nucleon as given in
eq. (\ref{lagrho}), manifest themselves in two spectral functions which
must be calculated from the two amplitudes $f^1_-$ and $f^1_+$ by
using eq. (\ref{ampfrho}), but how to assign the
propagators (spectral functions) is not yet clear. We could either
define (a) separate spectral functions for the amplitudes $A$ and $B$ or
(b) spectral functions for the vector and tensor part of the coupling.

\begin{itemize}
\item[a)]
By using the {\em ansatz }
\begin{eqnarray}
A^{(-)}_{\rho}&=&g_{\rho\pi\pi}\frac{g_t}{m_N}p_t q_t x
D^A_{\rho}(t) \nonumber \\
B^{(-)}_{\rho}&=&-g_{\rho\pi\pi}(g_v+g_t)
D^B_{\rho}(t),
\end{eqnarray}
the first method leads us to the spectral functions
\begin{eqnarray}
\rho^A_{\rho}(t)&=&-\frac{12\pi}{p_t^2} \frac{m_N}{g_{\rho\pi\pi}g_t}
\left(\frac{m_N}{\sqrt{2}}Im(f^1_-(t))-Im(f_+^1(t))\right) \nonumber \\
\rho^B_{\rho}(t)&=&\frac{12
\pi}{\sqrt{2}}\frac{Im(f_-^1(t))}{g_{\rho\pi\pi}(g_v+g_t)},
\end{eqnarray}
which we now use to construct the amplitude for correlated $\pi\pi$ exchange
via
\begin{eqnarray}
A^{(-)}_{2\pi}&=&\frac{1}{\pi}\int dm_{\rho}^2
\rho^A_{\rho}(m^2_{\rho})A^{(-)}_{\rho}\nonumber \\
B^{(-)}_{2\pi}&=&\frac{1}{\pi}\int dm_{\rho}^2
\rho^B_{\rho}(m^2_{\rho})B^{(-)}_{\rho}.
\end{eqnarray}
We finally obtain for the amplitudes
\begin{eqnarray}\label{abone}
A_{2\pi}^{(-)}&=&-6(\frac{1}{2}Q^{\mu}(p_1+p_3)_{\mu}) \int dt'
 \frac{1}{p_{t'}^2}
\left(\sqrt{2}m_N Im(f_-^1(t'))-2Im(f_+^1(t'))\right) P_{\rho}(t,t')
 \nonumber \\
B_{2\pi}^{(-)}&=&-6\sqrt{2} \int dt'  Im(f_-^1(t'))P_{\rho}(t,t').
\end{eqnarray}

\item[b)]The different momentum dependence of the vector and tensor coupling as given 
in eq. (\ref{lagrho}) may lead to different dressing of these couplings
in expressing the amplitude by a dressed propagator (as displayed in Fig. \ref{figpic}).
The second possibility of defining spectral functions for the vector
and tensor coupling takes this into account and leads us to a system of coupled equations
\begin{eqnarray}
-g_{\rho\pi\pi}(g_vD_{\rho}^v(t)+g_tD^t_{\rho}(t))&=&\frac{12\pi}{\sqrt{2}}f^1_-(t)
\nonumber \\
g_{\rho\pi\pi}\frac{g_t}{m_N} p_t q_t x D^t_{\rho}(t)&=& \frac{12\pi}{p_t^2}
p_t q_t x
\left(\frac{m_N}{\sqrt{2}}f^1_-(t)-f^1_+(t)\right).
\end{eqnarray}
Solving this system of popagators and using the definiton of the
spectral function given in eq. (\ref{propspec}) as well as the relations
\begin{eqnarray}
A^{(-)}_{2\pi}&=&\frac{1}{\pi}\int dm_{\rho}^2
\rho^t_{\rho}(m^2_{\rho})A^{(-)}_{\rho} \nonumber \\
B^{(-)}_{2\pi}&=&-\frac{g_{\rho\pi\pi}}{\pi}\int dm_{\rho}^2
(\rho^v_{\rho}(m^2_{\rho})g_v+\rho^t_{\rho}(m^2_{\rho})g_t)
P_{\rho}(t,m^2_{\rho}).
\end{eqnarray}
for integrating over the $\rho$ mass distribution we get
\begin{eqnarray}\label{abtwo}
A^{(-)}_{2\pi}&=&12\frac{\frac{1}{2}Q^{\mu}(p_1+p_3)_{\mu}}{m_N} \int
dt'\frac{m_N}{p^2_{t'}}
\left(Im(f^1_+(t'))-\frac{m_N}{\sqrt{2}}Im(f^1_-(t'))\right)P_{\rho}(t,t')
\nonumber \\
&=&12\frac{\frac{1}{2}Q^{\mu}(p_1+p_3)_{\mu}}{m_N} \int
dt'Im(\Gamma_2(t'))P_{\rho}(t,t') \nonumber \\
B^{(-)}_{2\pi}&=&12\int dt' \frac{m_N}{p^2_{t'}}
\left(Im(f^1_+(t'))-\frac{t'}{4\sqrt{2}m_N}Im(f^1_-(t'))\right)P_{\rho}(t,t')\nonumber
\\
&&-12\int dt' \frac{m_N}{p^2_{t'}}
\left(Im(f^1_+(t'))-\frac{m_N}{\sqrt{2}}Im(f^1_-(t'))\right)P_{\rho}(t,t')
\nonumber \\
&=&-12\int dt' Im(\Gamma_2(t')+\Gamma_1(t'))P_{\rho}(t,t'),
\end{eqnarray}
where we have written our result also in terms of the $\Gamma$
amplitudes defined in eq. (\ref{gammaamp})

The amplitude $B^{(-)}_{2\pi}$ can be reduced further by using
$p_{t'}^2=t'/4-m_N^2$ to give
\begin{equation}\label{finalb}
B^{(-)}_{2\pi}=-12\int dt'\frac{Im(f^1_-(t'))}{\sqrt{2}}P_{\rho}(t,t').
\end{equation}
\end{itemize}

As can be seen by comparing $A^{(-)}_{2\pi}$ from eq. (\ref{abtwo}) and
$B^{(-)}_{2\pi}$ from eq. (\ref{finalb}) with the amplitudes in eq. (\ref{abone})
both methods lead to the same amplitudes.
Therefore it does not matter how
we assign the propagators to the different amplitudes or coupling
schemes; they all lead to the same result. This makes our prescription for
calculating the correlated $\pi\pi$ exchange in the $\rho$ channel
{\it unambiguous}.

In comparing the amplitudes from the dispersion relation approach and
the amplitudes from our approach using the spectral functions, we see,
that the amplitudes of eq. (\ref{dispgamma}) generated by the dispersion
over the $\Gamma$ amplitudes are in complete agreement with our
amplitudes from eq. (\ref{abtwo}).  Furthermore we recognize, that the only
difference between our amplitudes from eq. (\ref{abone}) and the amplitudes
from dispersing the $f$ amplitudes of eq. (\ref{dispf}) is the position of
the term $\frac{1}{p_t^2}$. In our case this term appears inside the
integration whereas in eq. (\ref{dispf}) it appears in front of the
integral. A dispersion relation over the amplitudes $\frac{f}{p_t^2}$
would therefore lead also to the eq. (\ref{dispgamma}). The
dispersion relation over $\frac{f}{p_t^2}$ was done for the $\sigma$
exchange right from the beginning, since the 
known (model dependent) part of the $f^0_+$ amplitude does not 
have the necessary asymptotic behavior as $t \to \infty$ \cite{Frazer60}.

\section{Summary}

To summarize, we have introduced an alternative method for
the construction of amplitudes for the correlated $\pi\pi$ exchange that
does not suffer from ambiguities of the $\rho$ exchange in approaches through
dispersion relations. In the $\sigma$ channel our result
agrees completely with the on-shell results from dispersion theory, however
by starting from a field theoretically constructed potential, 
our approach removes ambiguities in the construction of the off shell amplitudes.

In the $\rho$ channel the source of ambiguity in the coupling constants is
traced to the choice whether to perform a dispersion relation over the
amplitudes $f$ or the ratio $\frac{f}{p_t^2}$. The latter choice agrees
completely with dispersion relations over the $\Gamma$ amplitudes used by
H\"ohler and Pietarinen.
Our approach solves the problem of non-uniqueness and yields
amplitudes that are in complete agreement with
the results from dispersion relations over the $\Gamma$ amplitudes
for on-shell processes.

\acknowledgments
We are grateful to I.R. Afnan and G. H\"ohler for stimulating
discussions.  Furthermore we would like to thank J.W. Durso for
carefully reading the manuscript.  Two of us (O.K. and J.S.) would
like to thank A.W. Thomas for the hospitality they enjoyed during
their stay at the {\it Special Research Centre for the Structure of
  Subatomic Matter} in Adelaide/Australia, where part of this work was
done.  O.K. thanks the Deutsche Forschungsgemeinschaft for financial
support under Grant No. 447 AUS-113/3/0 during his stay in Adelaide.
C.H. acknowledges support from the Alexander-von-Humboldt foundation
as a Theodor-Lynen fellow. This work was supported in part by the U.S.
Department of Energy under Grant No. DE-FG03-97ER41014.

\appendix

\section{Decomposition of the transfer matrix}\label{zerleg}

\begin{figure}
\begin{center}
\epsfig{file=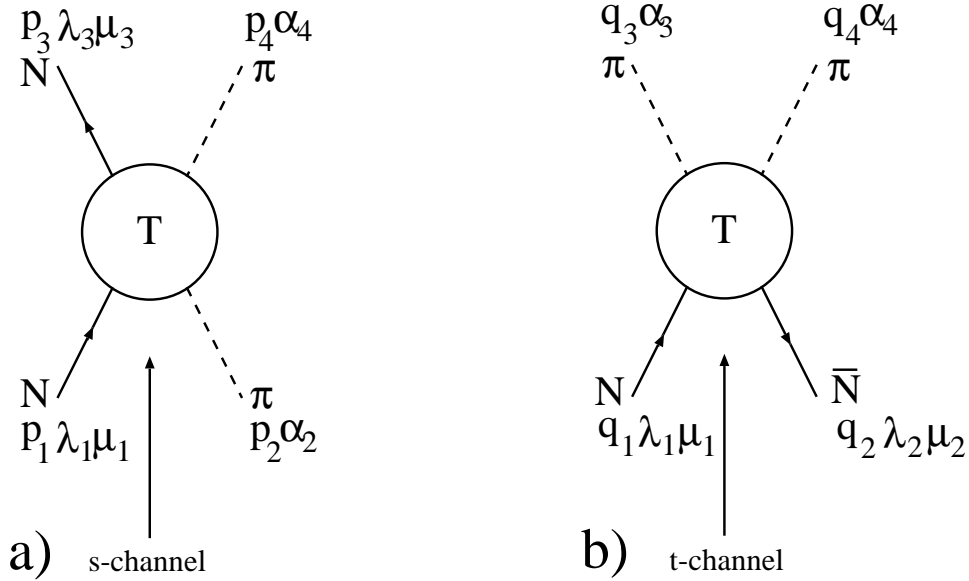,height=8cm}
\caption{Notation in the reaction channels $\pi N\to \pi N$ a) and
$N\bar{N}\to \pi\pi$ b)}
\label{fignot}
\end{center}
\end{figure}

The transfer matrix $T$ is related to the standard $S$-matrix by
\begin{displaymath}
S_{fi}=\delta_{fi}-i(2\pi)^{-2}\delta^{(4)}(P_f-P_i)\left(\frac{m_N}{E_{p_1}}\frac{m_N}{E_{p_3}}\right)^{\frac{1}{2}}(2\omega_{p_2}2\omega_{p_4})^{-\frac{1}{2}}
T_{fi}.
\end{displaymath}
In the $s$-channel the transfer matrix $T$ can be represented as
\begin{equation}\label{zerlegs}
T_s(p_3,p_4,p_1,p_2)=
\bar{u}(\vec{p}_3,\lambda_3)\xi^{\dagger}(\mu_3)\zeta^{\dagger}(\alpha_4)
\hat{T}(s,t)
u(\vec{p}_1,\lambda_1)\xi(\mu_1)\zeta(\alpha_2),
\end{equation}
where $u(\vec{p},\lambda)$ is the Dirac spinor of the nucleon with
momentum $\vec{p}$ and helicity $\lambda$.
The isospin wave functions -- $\xi$ for the nucleon and $\zeta$ for the
pion -- explicitly depend on the third components as given in Fig.
\ref{fignot}.

The operator $\hat{T}$ acts in spin-momentum and isospin space and can be
decomposed into
\begin{equation}
\hat{T}(s,t)=\hat{T}^{(+)}(s,t){\bf
1}-\hat{T}^{(-)}(s,t)\vec{\tau}\cdot\vec{t},
\end{equation}
where $\vec{\tau}$ and $\vec{t}$ are the isospin operators of nucleon
and pion, respectively. The index $(+)=$ symmetric and
$(-)= $antisymmetric tells us how the amplitudes behave under
exchange of the two pions.
These amplitudes can be represented by
\begin{equation}\label{abs}
\hat{T}^{(\pm)}=-(A^{(\pm)}(s,t)I_4+\gamma^{\mu}Q_{\mu} B^{(\pm)}(s,t)),
\end{equation}
where $Q=\frac{1}{2}(p_2+p_4)$.

The $T$ matrix in the $t$-channel can be decomposed in the same way
\begin{equation}\label{zerlegt}
T_t(q_3,q_4,q_1,q_2)=
\bar{v}(\vec{q}_2,\lambda_2)\xi^{\dagger}(\mu_2)\zeta^{\dagger}(\alpha_3)
\hat{T}(s,t)
u(\vec{q}_1,\lambda_1)\xi(\mu_1)\zeta(\alpha_4),
\end{equation}
where the amplitudes $\hat{T}$ and therefore also $A^{(\pm)}$ and
$B^{(\pm)}$ are the same functions as in eq. (\ref{zerlegs}) and
eq. (\ref{abs}), however, in a different kinematic domain.

In order to isolate the contributions of the $\sigma$ and $\rho$ we
have to perform a partial wave decomposition of the amplitudes
$A^{(\pm)}$ and $B^{(\pm)}$,
\begin{equation}\label{pwia}
A^{(\pm)}(s,t)=\sum_{J} \frac{1}{2}(2J+1)P_{J}(x)A^{(\pm)}(t),
\end{equation}
and the same for $B^{(\pm)}$. The argument $x$ of the
Legendre polynomials $P_J(x)$ is the cosine of the scattering angle in
the $t$-channel.  Due to crossing symmetry, the amplitudes $A^{(-)}_J$
and $B^{(+)}_J$ have to vanish for even $J$ and $A^{(+)}_J$ and
$B^{(-)}_J$ have to vanish for odd $J$, so that the index $J$ determines
also the symmetry $(\pm)$, which can then be dropped from the
partial wave amplitudes.

The $f$ amplitudes introduced by Frazer and Fulco \cite{Frazer60} are
free of kinematical singularities. They are connected to the partial
wave amplitudes in the following way:
\begin{eqnarray} \label{famp}
f^J_+(t)&=&\frac{1}{8\pi}\left(\frac{-p_t^2}{(p_tq_t)^J}A_J+\frac{m_N((J+1)B_{J+1}+JB_{J-1})}{(2J+1)(p_tq_t)^{J-1}}\right)
\nonumber \\
f^J_-(t)&=&\frac{1}{8\pi}\frac{\sqrt{J(J+1)}}{2J+1}\frac{1}{(p_tq_t)^{J-1}}(B_{J-1}-B_{J+1}).
\end{eqnarray}
The momenta 
\begin{eqnarray}
p_t&=&\sqrt{\frac{t}{4}-m_N^2} \quad \mbox{and} \\
q_t&=&\sqrt{\frac{t}{4}-m_{\pi}^2}
\end{eqnarray}
 are the $t$-channel momenta of the
$N\bar{N}$ and $\pi\pi$ system in their c.m., respectively.
The index $\pm$ gives the helicity of the nucleon $\pm\frac{1}{2}$. The
helicity of the antinucleon is fixed to $+\frac{1}{2}$.
Using these amplitudes, the invariant amplitudes can be written as
\begin{eqnarray} \label{aampfamp}
A^{(\pm)}(s,t)&=&\frac{8\pi}{p^2_t}\sum_{J}\frac{1}{2}(2J+1)(p_tq_t)^J
\times\nonumber \\
&&\vspace{3cm}\times\left(\frac{m_N}{\sqrt{J(J+1)}}xP'_J(x)f^J_-(t)-P_J(x)f^J_+(t)\right)
\nonumber \\
B^{(\pm)}(s,t)&=&8\pi\sum_{J}\frac{1}{2}(2J+1)\frac{(p_tq_t)^{J-1}}{\sqrt{J(J+1)}}P'_J(x)f^J_-(t),
\end{eqnarray}
where $P'_J(x)=\frac{d}{dx}P_J(x)$. The symmetry $(\pm)$ of the
amplitudes is determined by summing only even (odd) $J$. The $\sigma$
and $\rho$ contribution is contained in the $J=0$ and $J=1$ term in
eq. (\ref{aampfamp}), respectively. Explicitly, the amplitudes for the
$\sigma$ channel read
\begin{eqnarray}\label{ampfsigma}
A^{(+)}_{\sigma}(s,t)&=&-\frac{4\pi}{p_t^2}f^0_+(t) \nonumber \\
B^{(+)}_{\sigma}(s,t)&=&0,
\end{eqnarray}
and for the $\rho$ channel
\begin{eqnarray}\label{ampfrho}
A^{(+)}_{\rho}(s,t)&=&12\pi\frac{p_tq_tx}{p^2_t}\left(\frac{m_N}{\sqrt{2}}f^1_-(t)-f^1_+(t)\right)
\nonumber \\
B^{(+)}_{\rho}(s,t)&=&6\sqrt{2}\pi f^1_-(t).
\end{eqnarray}

\end{document}